\documentclass[aps,prd,nofootinbib]{revtex4}
\usepackage{multirow}
\usepackage[english]{babel}
\usepackage{amsmath}
\usepackage{amssymb}
\usepackage{epsfig}
\usepackage{graphics,psfrag,rotating}
\usepackage{graphicx}% Include figure files
\usepackage{dcolumn}% Align table columns on decimal point
\usepackage{bm}% bold math
\begin{document}
\newcommand{\Od}{{\cal O}}
\newcommand{\lsim}   {\mathrel{\mathop{\kern 0pt \rlap
  {\raise.2ex\hbox{$<$}}}
  \lower.9ex\hbox{\kern-.190em $\sim$}}}
\newcommand{\gsim}   {\mathrel{\mathop{\kern 0pt \rlap
  {\raise.2ex\hbox{$>$}}}
  \lower.9ex\hbox{\kern-.190em $\sim$}}}

%\preprint{APS/123-QED}

\title{Detection of branon dark matter with gamma ray telescopes} %% Paper title goes here

\author{J. A. R. Cembranos$^{1}\footnote{E-mail: cembra@fis.ucm.es}$, A. de la Cruz-Dombriz$^{2,3}\footnote{E-mail: alvaro.delacruzdombriz@uct.ac.za}$, V. Gammaldi$^{1}\footnote{E-mail: vivigamm@pas.ucm.es}$ and A.\,L.\,Maroto$^{1}\footnote{E-mail: maroto@fis.ucm.es}$}
\address{$^{1}$Departamento de  F\'{\i}sica Te\'orica I, Universidad Complutense de Madrid, E-28040 Madrid, Spain}%
\address{$^{2}$  Astrophysics, Cosmology and Gravity Centre (ACGC), University of Cape Town, Rondebosch, 7701, South Africa}
\address{$^{3}$ Department of Mathematics and Applied Mathematics, University of Cape Town, 7701 Rondebosch, Cape Town, South Africa}
%\affiliation{Departamento de  F\'{\i}sica Te\'orica I, Universidad Complutense de Madrid, E-28040 Madrid, Spain}%

%\affiliation{...Minessota}
%\affiliation{INFN, sezione di Torino, via P. Giuria 1, 10125 Torino, Italy}
\date{\today}% It is always \today, today,
             %  but any date may be explicitly specified.

\begin{abstract}
Branons are new degrees of freedom that appear in flexible brane-world models
corresponding to brane fluctuations. These new fields can behave as standard
weakly interacting massive particles (WIMPs) with a %and have associated an
significant associated thermal relic density. We analyze the present constraints from their spontaneous annihilations
into photons for EGRET, Fermi-LAT and MAGIC, and the prospects for detection in future
Cherenkov telescopes. In particular, we focus on possible signals coming from
the Galactic Center and different dwarf spheroidals, such as Draco, Sagittarius, Canis Major
and SEGUE 1. We conclude that for those targets, present observations are below
the sensitivity limits for branon detection by assuming standard dark matter distributions
and no additional boost factors.
However, future experiments such as CTA
could be able to detect gamma-ray photons coming  from the annihilation of branons with masses
higher than 150 GeV.
\end{abstract}

%11.10.Kk   Field theories in dimensions other than four
%12.60.-i   Models beyond the standard model
%95.35.+d   Dark matter
%98.80.Cq   Particle-theory and field-theory models of the early
%              Universe (including cosmic pancakes, cosmic strings,
%              chaotic phenomena, inflationary universe, etc.)

%\maketitle must follow title, authors, abstract and pacs
\maketitle

%\thispagestyle{fancy}


\section{Introduction}
\label{aba:sec1}

Dark matter (DM) is one of the most intriguing puzzles in physics.
The fact that DM cannot be made of any of the known
particles is one of the most appealing arguments for the existence of new physics.
The experimental search for its nature needs
the interplay of new collider experiments \cite{colliders} and
astrophysical observations. These last ones used to be classified
in direct or indirect searches (although there are other alternatives
\cite{Feng:2003xh, structureformation}): Elastic scattering of DM particles
from nuclei should lead directly to observable nuclear recoil signatures.
On the other hand, DM might be detected indirectly, by observing
the products of their annihilation into Standard Model (SM) particles.
We will focus our discussion on this last alternative.

It has been found that massive brane fluctuations (branons)
are interesting candidates for DM in brane-world models
with low tension \cite{CDM}. From the point of view of the 4-dimensional
effective phenomenology, massive branons are new
pseudo-scalar fields which can be understood as the
pseudo-Goldstone bosons corresponding to the spontaneous
breaking of translational
 invariance in the bulk space produced by the presence of the
brane \cite{BR, ACDM}. They are prevented from decaying into SM
particles by  parity invariance on the brane.
The  SM-branon low-energy effective Lagrangian
\cite{BR, ACDM}  can be written as
\begin{eqnarray}
{\mathcal L}_{Br}\,=\,
\frac{1}{2}g^{\mu\nu}\partial_{\mu}\pi^\alpha
\partial_{\nu}\pi^\alpha-\frac{1}{2}M^2\pi^\alpha\pi^\alpha
%\nonumber+  \\
%&&
+
\frac{1}{8f^4}\left(4\partial_{\mu}\pi^\alpha
\partial_{\nu}\pi^\alpha-M^2\pi^\alpha\pi^\alpha g_{\mu\nu}\right)
T^{\mu\nu}
%_{SM}
\,\label{lag}
\end{eqnarray}
where $\alpha=1\dots N$, with $N$ the number of branon
species.

One can see that branons interact by pairs with the SM
energy-momentum tensor $T^{\mu\nu}$ and that the coupling
is suppressed by the brane tension $f^4$.
Limits on the model parameter from tree-level processes in colliders
are given by %briefly summarized in Table \ref{tabHad}, where not
%only
present restrictions coming from HERA, Tevatron and LEP-II, % are provided,
but also  prospects for future colliders such as ILC, LHC or CLIC
can be found in \cite{ACDM,L3,CrSt,Rad}.
Additional bounds from astrophysics and
cosmology were obtained in \cite{CDM}.

Even if branons are stable, two of them may annihilate into ordinary particles such as quarks, leptons and gauge bosons. Their annihilation in different astrophysical objects (galactic halo, Sun, Earth, etc.) produces cosmic rays to be discriminated through distinctive signatures from the background. After branon annihilation, a cascade process would occur and in the end, the particle species that can be potentially observed would be neutrinos, gamma rays, positrons and antimatter (antiprotons, antihelium, antideuterons, etc.) that might be detectable by means of different experimental devices. Among them, neutrinos and gamma rays have the advantage of maintaining their original direction of motion. On the contrary, charged antimatter searches are hindered by the modification of the propagation trajectories.

The paper has been arranged as follows: in Section II we give a general overview for dark matter indirect detection by using gamma rays and we present  the thermal averaged cross sections for branon scenarios. In Section III, we summarize the most relevant gamma-ray telescopes operating nowadays, both satellite and ground-based experiments. Section IV is then devoted to the calculation of the minimum detectable fluxes from different targets by using the already mentioned detectors. Finally,  Section V contains the main conclusions about the performed analyses.

\section{Gamma rays}

The differential gamma ray flux from annihilating DM particles in galactic sources can be written as \cite{BUB,FMW,Ce10}:
\begin{eqnarray}
\frac{\text{d}\,\Phi_{\gamma}^{\text{DM}}}{\text{d}\,E_{\gamma}} =
\frac{1}{4\pi M^2}\sum_i\langle\sigma_i v\rangle
\frac{\text{d}\,N_\gamma^i}{\text{d}\,E_{\gamma}}\, \times\, \frac{1}{\Delta\Omega}\int_{\Delta\Omega}\text{d}\Omega\int_{l.o.s.} \rho^2[(s)] \text{d}s
\label{flux}
\end{eqnarray}
where the second term on the r.h.s. of this equation represents the astrophysical factor, i.e, the integral of the DM mass density
profile, $\rho(r)$, along the path (line of sight, $l.o.s.$) between the source and
the gamma ray detector divided by the detector solid angle.
On the other hand, the first term is the particle dependent part, with
$\langle\sigma_i v\rangle$ the thermal averaged
annihilation cross-section of two DM particles into two SM
particles (labeled by the subindex $i$). The number of photons produced in each decaying channel
per energy interval
$\text{d}\,N_\gamma^i/\text{d}\,E_{\gamma}$
involves decays and/or hadronization of unstable products such as quarks and leptons.
Due to the non-perturbative QCD effects, the analytical
calculation of these decay chains is a hard task to be accomplished
and therefore requires Monte Carlo events generators
such as PYTHIA \cite{PYTHIA} particle physics software.
By using this software and after exhaustive and statistically significant simulations, the gamma-ray spectra
were obtained in \cite{Ce10}. In these works, three different parametrizations were found to be able to fit
all the available data in the whole accessible photon
energy range for the studied channels:
the first one for leptons and quarks (except for the top) is given by
\begin{eqnarray}
x^{1.5}\frac{\text{d}N_{\gamma}}{\text{d}x}\,=\, a_{1}\text{exp}\left(-b_{1} x^{n_1}-b_2 x^{n_2} -\frac{c_{1}}{x^{d_1}}+\frac{c_2}{x^{d_2}}\right) + q\,x^{1.5}\,\text{ln}\left[p(1-x)\right]\frac{x^2-2x+2}{x}\,;
\label{general_formula}
\end{eqnarray}
the second for the top quark is encapsulated in the following expression
\begin{eqnarray}
x^{1.5}\frac{\text{d}N_{\gamma}}{\text{d}x}\,=\, a_{1}\,\text{exp}\left(-b_{1}\, x^{n_1}-\frac{c_{1}}{x^{d_1}}-\frac{c_{2}}{x^{d_2}}\right)\left\{\frac{\text{ln}[p(1-x^{l})]}{\text{ln}\,p}\right\}^{q}\,;
\label{general_formula_t}
\end{eqnarray}
and the last one for the $W$ and $Z$ bosons
\begin{eqnarray}
x^{1.5}\frac{\text{d}N_{\gamma}}{\text{d}x}\,=\, a_{1}\,\text{exp}\left(-b_{1}\, x^{n_1}-\frac{c_{1}}{x^{d_1}}\right)\left\{\frac{\text{ln}[p(j-x)]}{\text{ln}\,p}\right\}^{q}.
\label{general_formula_W_Z}
\end{eqnarray}
The variable $x$ stands for $x\equiv E_{\gamma}/M$ where $E_{\gamma}$ is the photon energy and $M$ holds for the DM candidate mass.
The concrete values for the constants in the above expressions (3)-(5)
are given in \cite{Ce10} and are also available online in numerical codes \cite{Online}. The contributions
of decreasing and increasing exponential factors change depending upon the channel. The asymptotic logarithmic terms
appearing in expressions (3)-(5) %, for $x$ close to one,
are a consequence of the Weizsacker-Williams effect \cite{cutoff}, although
with either additive or multiplicative behavior depending on the annihilation channel. 
For each channel some parameters are WIMP mass dependent whereas the remaining ones are constants.

\begin{figure}[bt]
\begin{center}
\epsfxsize=13cm
\resizebox{11cm}{8cm}
%{8.8cm}{6.4cm}
{\includegraphics{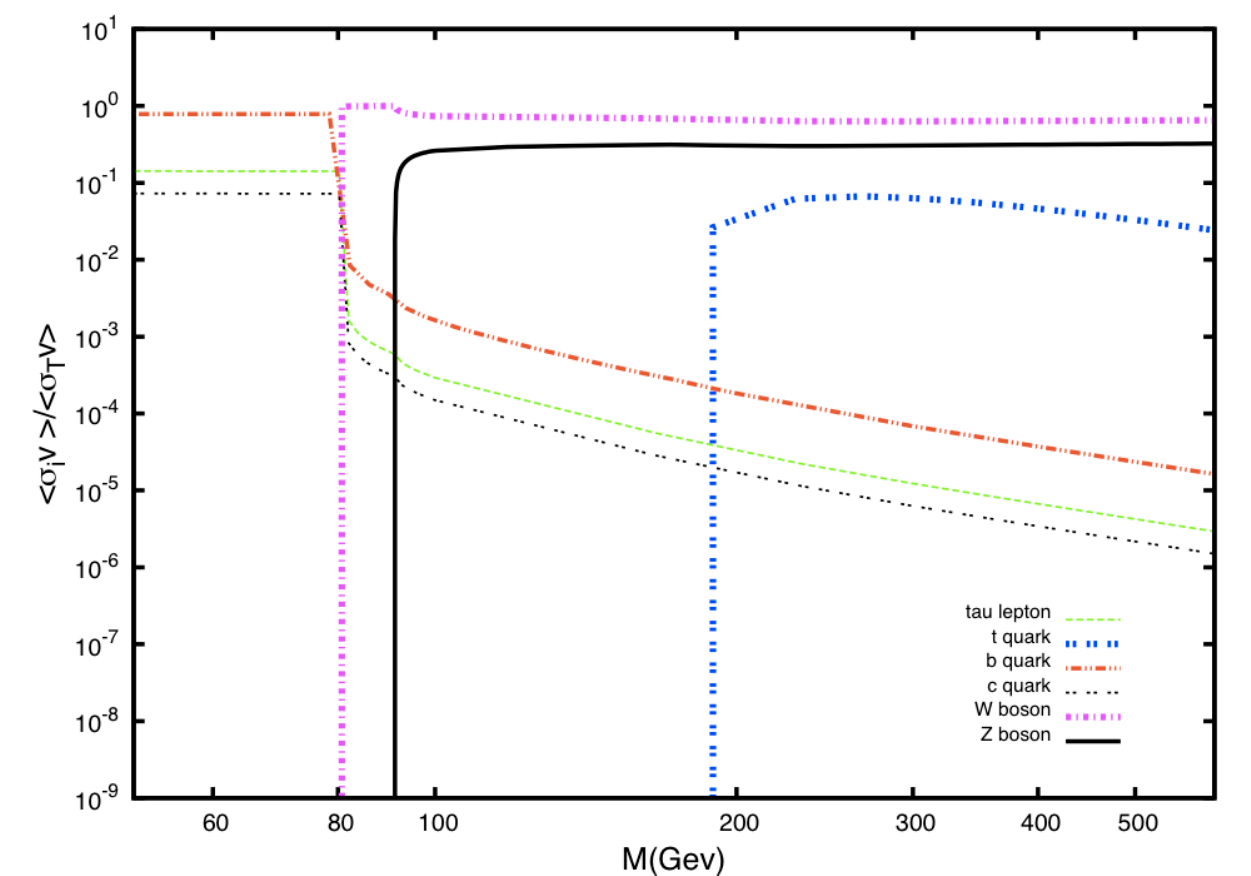}}
\caption {\footnotesize{Branon annihilation branching ratios into SM particles. In the case of heavy branons the main contribution to the photon flux comes from branon annihilation into $W^+W^-$ and $ZZ$. The contribution from heavy fermions, i.e. annihilation into top-antitop can be shown to be subdominant. On the contrary, if $M < m_{W,\,Z} $, the annihilation into $W$ or $Z$ bosons is kinematically forbidden and it is necessary to take into account the rest of channels, mainly annihilation into the heaviest possible quarks. } }
\label{BR}
\end{center}
\end{figure}

In the case of heavy branons, the main contribution to the photon flux comes from branon annihilation into
$ZZ$ and $W^+ W^-$ (Fig. \ref{BR}). The contribution from heavy fermions, i.e., annihilation
into top-antitop can be shown to be subdominant \cite{indirect}.
In this case, the produced high-energy gamma photons could be in the range
(30 GeV-10 TeV), detectable by Atmospheric
Cherenkov Telescopes (ACTs)  such as MAGIC \cite{Mag, Mag11}.

On the contrary, if $M<m_{W,Z}$, the
annihilation into $W$ or $Z$ bosons is kinematically forbidden and it is
necessary to take into account the rest of  channels, mainly
annihilation into the heaviest possible quarks \cite{Silk} as can be seen in Fig. \ref{BR}.
In this case, the photon fluxes would be in the range detectable by
space-based gamma ray observatories \cite{indirect}
such as EGRET \cite{EGRET} and FERMI \cite{Fer, FER},
with better sensitivities around 30 MeV-300 GeV.

Therefore, in order to compute the 
final spectrum coming from branon annihilation, we need to know the total annihilation cross section and its annihilation branching ratios into SM particles. For branons, the annihilation cross sections only depend on the spin and mass of the particle, and the   expressions for branons thermal average annihilations $\langle \sigma_{i} v\rangle$ in different SM particles channels, i.e., in Dirac fermions, massive gauge field, massless gauge field and complex scalar field were calculated in \cite{indirect}. For instance, the leading term for non-relativistic branons decaying into a Dirac fermion $\psi$ with mass $m_\psi$, is given by:
\begin{eqnarray}
\langle \sigma_{\psi} v\rangle\,=\,\frac{1}{16\pi^2f^8}M^2 m_\psi^2\left(M^2-m_\psi^2\right)\,\sqrt{1-\frac{m_\psi^2}{M^2}}\,;
\end{eqnarray}
For a massive gauge field $Z$, of mass $m_Z$, it reads:
\begin{eqnarray}
\langle \sigma_{Z} v\rangle\,=\,
\frac{1}{64\pi^2f^8}
M^2\,\left( 4\,M^4 - 4\,M^2\,{m_Z}^2 + 3\,
{m_Z}^4 \right)\,{\sqrt{1 - \frac{{m_Z}^2}{M^2}}}
\,;
\end{eqnarray}
whereas for a massless gauge field $\gamma$, the leading order is zero:
\begin{eqnarray}
\langle \sigma_{\gamma} v\rangle&=&0;
\end{eqnarray}
and, finally, for a (complex) scalar field $\Phi$ with mass $m_\Phi$:
\begin{eqnarray}
\langle \sigma_{\Phi} v\rangle\,=\,\frac{1}{32\pi^2f^8}
M^2\,{\left( 2\,M^2 + {m_\Phi}^2 \right) }^2\,
{\sqrt{1 - \frac{{m_\Phi}^2}{M^2}}}
\,.
\end{eqnarray}

It is worth noting that there would be a gamma ray line from direct annihilation into photons since branons couple directly to them, producing a monochromatic signal at the energy equal to the branon mass. However, this annihilation takes place in $d$-wave channel and it is highly suppressed.

\section{Gamma-ray telescopes}

For gamma-ray detection several devices have been used in the last years to observe galactic and extragalactic sources. Here we focus on both satellite and ground-based detectors of gamma-ray photons.

\subsection{Satellite experiments}
 We will analyze the performance of two satellite telescopes: EGRET and Fermi-LAT.
\subsubsection{$\text{EGRET}$}
EGRET (Energetic Gamma-Ray Experiment Telescope) has detected gamma rays in the interval 0.02 to 30 GeV. This telescope was carefully calibrated at SLAC with a quasi mono-energetic beam in the energy range of 0.02 to 10 GeV. Although the energy range was extended up to higher energies (30 GeV) by using Monte Carlo simulations, we have used only data below 10 GeV \cite{EGRET} due to the associated larger uncertainty at high energy. The angular resolution was energy dependent, with a 67$\%$ confinement angle of 5.5$^\circ$ at 100 MeV, falling to 0.5$^\circ$ at 5 GeV on axis. The energy resolution of EGRET was 20-25$\%$ over most of its range of sensitivity. Absolute arrival times for photons were recorded with approximately 50 $\mu \text{s}$ accuracy. The overall normalization error is usually quoted as 15 $\%$ \cite{EGRET}.

\subsubsection{$\text{FERMI}$}
The Large Area Telescope (LAT) is the principal scientific instrument on the FERMI Gamma Ray Space Telescope spacecraft, originally called the Gamma-Ray Large Area Space Telescope (GLAST). The LAT is an imaging high-energy gamma ray telescope covering the energy range from about 20 MeV to more than 300 GeV \cite{Fer}. The LAT's field of view covers about $20\%$ of the sky at any time, and it scans continuously, covering the whole sky every three hours. The LAT measures the tracks of the electron and positron that result when an incident gamma ray undergoes
pair-conversion, preferentially in a thin, high-Z foil, and measures the energy of the subsequent electromagnetic shower
that develops in the telescope calorimeter. The development of the reconstruction relies heavily on the Monte Carlo simulation of the
events. The background model for this device includes cosmic rays and Earth's albedo gamma rays within the energy range 10 MeV to 1 GeV. Particles that might either make non astrophysical gamma rays and/or need to be rejected as background are included. The model does not include X-rays nor soft gamma rays that might cause individual detectors within the LAT to be activated \cite{Fer,FER}. Recent investigations \cite{Nieto:2011sx} have looked for potential DM subhalos targets for ACTs and FERMI at energies higher than 100 GeV claiming that multi-wavelength observations %by using FERMI and CTAs
could play an essential role in DM indirect searches.

\subsection{Ground-based experiments}

On the other hand, the measurement of very high-energy gamma rays using very large ground-based Cherenkov telescopes is a recent addition to the variety of experiments at the interface between particle physics, astrophysics and cosmology. 

\subsubsection{MAGIC}
In particular, the MAGIC experiment for ground-based gamma ray astronomy is the largest of the third generation Cherenkov telescope experiments, located at the La Palma (Canary Islands) observatory, 2200 m above sea level. It consists of a system of two telescopes operating in a stereoscopic mode. The stereoscopic system allows for improved background rejection specially at low energies (and therefore high sensitivity), improved energy and angular resolution, and a low energy threshold. The initial setup (MAGIC-I) consists of a single telescope (focal length of 17 m). The threshold for gamma detection is around 60-70 GeV with classical photomultiplier tubes (PMTs); future high-quantum efficiency red-extended PMTs are expected to achieve a lower threshold. MAGIC has the best light collection that has been attempted so far: the largest mirror with an active surface of 234 $\text{m}^2$, combined with the best available photomultiplier tubes that can be obtained, of a quantum efficiency around 30\%. As a result, MAGIC is more sensitive to electromagnetic showers of lower energy, and does much to close the gap existing between satellite gamma ray detectors (that can go up to 10 GeV energy) and Cherenkov telescopes (that presently start at energies higher than $100$ GeV). MAGIC-I has a threshold trigger energy of $\sim50$ GeV, and an analysis threshold of $\sim70$ GeV at small zenith angle, which also permits to observe sources with higher redshift than in the past \cite{Mag11,Mag}.
%With the standard MAGIC-I analysis and reconstruction software, after the calibration and image cleaning, the hadronic background rejection is achieved through a multivariate method called Random Forest %(RF)
\subsubsection{CTA}
We are also interested in estimating the prospects for future experiments as the Cherenkov Telescope Array (CTA) project, that is an initiative to build the next generation of ground-based very high energy gamma-ray instruments. Current systems of Cherenkov telescopes use at most four telescopes, providing best stereo-imaging of particle cascades over a very limited area, with most cascades viewed by only two or three telescopes. An array of many tens of telescopes will allow the detection of gamma-ray induced cascades over a large area on the ground, increasing the number of detected gamma rays dramatically, while at the same time providing a much larger number of views of each cascade. This results in both improved angular resolution and better suppression of cosmic-ray background events.  The CTA will explore our Universe in depth
at gamma-rays of Very High Energy (VHE), i.e.,
$E > 10$ GeV and investigate cosmic non-thermal processes, in close cooperation with observatories operating at other wavelength ranges of the electromagnetic spectrum, and those using other messengers such as cosmic rays and neutrinos. The design foresees a factor of $5-10$ improvement in sensitivity in the current very high energy gamma ray domain of about 100 GeV to some 10 TeV, and an extension of the accessible energy range from well below 100 GeV to above 100 TeV \cite{cta}.

%\begin{table}[htdp]
\begin{table}[t]
\begin{center}
\begin{tabular}{|c|c|c|c|c|}
\hline
\hline
\multicolumn{5}{|c|}{Draco dSph} \\
\hline
\hline
Experiment& EGRET & FERMI & MAGIC-I$ ^*$ & CTA \\
\hline
\multirow{2}{*}{$A_{Eff}$} & $10^3$(*) & \multirow{2}{*}{$10^4$} &\multirow{2}{*}{ $5\times10^8$} & \multirow{2}{*}{$10^{10}$} \\
 & $1.5\times10^3$ & & & \\
\hline
\multirow{2}{*}{$\Delta\Omega$} & \multirow{2}{*}{$10^{-3}$}  & $9\times10^{-5}$(*)  & \multirow{2}{*}{$10^{-5}$} & \multirow{2}{*}{$10^{-5}$} \\
 & & $10^{-5}$ & &  \\
\hline
$t_{exp}$ & \multicolumn{2}{|c|}{1\;yr} & 40 h & 50 h \\
\hline
\multirow{2}{*}{$\Phi_{Bg}$}  & \multicolumn{2}{|c|}{ $3.3\times10^{-7}$(*)} & \multirow{2}{*}{$1.9\times10^{-7}$(**)} & \multirow{2}{*}{$\sim10^{-7}$(**)} \\
&\multicolumn{2}{|c|}{ $6.7\times10^{-7}$} &   & \\
\hline
\multirow{2}{*}{$\Phi_{\gamma}^{(5)}$}  & $1.04\times10^{-6}$(*) & $1.14\times10^{-6}$(*)&  \multirow{2}{*}{$1.00\times10^{-7}$} & \multirow{2}{*}{$1.25\times10^{-8}$} \\
& $9.15\times10^{-7}$ & $8.55\times10^{-6}$ & & \\
\hline
\multirow{2}{*}{$\Phi_{\gamma}^{(2)}$} & $2.78\times10^{-7}$(*) & $2.97\times10^{-7}$(*)&  \multirow{2}{*}{$3.54\times10^{-8}$} & \multirow{2}{*}{$4.83\times10^{-9}$}  \\
& $2.84\times10^{-7}$ & $ 1.75\times10^{-6}$ & & \\
\hline
$\langle J\rangle_{\Delta\Omega}$ & 0.1 & \multicolumn{3}{|c|}{7.2} \\
\hline
\multirow{2}{*}{$N_\gamma^{(5)}\langle\sigma v\rangle/M^2$}  & $1.31\times10^5$(*) & $1.98\times10^3$(*) &  \multirow{2}{*}{$1.75\times10^2$} & \multirow{2}{*}{$21.8$} \\
& $1.15\times10^5$  &$1.49\times10^4$ & &\\
\hline
\multirow{2}{*}{$N_\gamma^{(2)}\langle\sigma v\rangle/M^2$} & $3.49\times10^4$(*) & $5.19\times10^2$(*) &  \multirow{2}{*}{$61.8$} & \multirow{2}{*}{$8.42$} \\
& $3.57\times10^4$ & $3.06\times10^3$ & & \\
\hline
\end{tabular}
\end{center}
\caption{
$\Phi_{\gamma}^{(5)}$ or $\Phi_{\gamma}^{(2)}$ (cm$^{-2}$s$^{-1}$sr$^{-1}$), where superscripts $(5)$ and $(2)$ denote the estimated minimum detectable flux at $5\sigma$ or $2\sigma$, for different detectors:
EGRET, FERMI, MAGIC-I and CTA associated with Draco.
The first three are current constraints, whereas the last one is a prospect for its sensitivity.
 $A_{Eff}$ (cm$^2$) denotes the typical effective area,
 $\Delta\Omega$ (sr) the angular acceptance,
 $t_{exp}$ (s) the exposure time and
  $\Phi_{Bg}$ (cm$^{-2}$s$^{-1}$sr$^{-1}$) the estimated total background flux.
In order to obtain $N_\gamma^{(2)}\langle\sigma v\rangle /M^2$ (cm$^3$s$^{-1}$),  we have used Eq. (\ref{flux}) with the astrophysical factor $\langle J\rangle_{\Delta\Omega}(10^{23}$GeV$^2$cm$^{-5}$sr$^{-1})$ given in \cite{Ev04} assuming a NFW profile for the DM distribution. Most part of the data are taken from \cite{Ev04}, except for the  values
marked with a single asterisk ($^*$) which are obtained from Ref. \cite{Berg08}.
The $\Phi_{Bg}$ signed with double asterisk ($^{**}$) is calculated by means of the Eq.(\ref{bg}) with $\epsilon=0.01$ %respectively
for both MAGIC-I and CTA projects \cite{Berg08,Ev04}.
}
\label{Drac}
\end{table}

%%%%%%

\section{Analysis and results}
Although there are other possibilities, as the Galactic center, the best targets to search for a DM annihilation signal seems to be dwarf spheroidals (dSphs), which are the smallest known systems dominated by DM. The astrophysical part $\langle J \rangle_{\Delta\Omega}$ of the gamma-ray flux (\ref{flux}) of each target depends upon the DM density. This factor is not very well known and introduces the most important uncertainties in these indirect detection analyses. A classic approach uses a Navarro-Frenk-White (NFW) profile. This profile is in good agreement with cold DM simulations and it allows an easy comparison with previous studies since it has been used by many authors. For this reason, we have assumed this profile for the Draco, Sagittarius and Canis Major dSphs and for the Galactic Center. In particular \cite{Ev04}:
\begin{equation}
\rho_{\text{NFW}}(r)=\frac{A}{r(r+r_s)^2}
\end{equation}
where $A$ is the overall normalization and $r_s$ the scale radius.

\begin{table}[t]
\begin{center}
\begin{tabular}{|c|c|c|c|}
\hline
\hline
\multicolumn{4}{|c|}{Sagittarius dSph} \\
\hline
\hline
Experiment& EGRET & FERMI & CTA \\
\hline
$A_{Eff}$ & $1.5\times10^3$ & $10^4$ & $10^{10}$ \\
\hline
$\Delta\Omega$ & $10^{-3}$  & $\sim10^{-5}$  & $10^{-5}$ \\
\hline
$t_{exp}$ & \multicolumn{2}{|c|}{1\;yr} & 100 h \\
\hline
$\Phi_{Bg}$ &  \multicolumn{2}{|c|}{$3.18\times10^{-6}$} & $2.7\times10^{-4}$(**) \\
\hline
$\Phi_{\gamma}^{(5)}$  & $1.59\times10^{-6}$ & $1.04\times10^{-5}$ & $1.25\times10^{-6}$ \\
\hline
$\Phi_{\gamma}^{(2)}$ & $5.62\times10^{-7}$ & $2.74\times10^{-6}$ & $4.83\times10^{-9}$ \\
\hline
$\langle J\rangle_{\Delta\Omega}$ & 1.3 & \multicolumn{2}{|c|}{36.9} \\
\hline
$N_\gamma^{(5)}\langle\sigma v\rangle/M^2$  & $1.53\times10^4$ & $3.53\times10^3$ & $4.25$ \\
\hline
$N_\gamma^{(2)}\langle\sigma v\rangle/M^2$ & $5.44\times10^3$ & $9.33\times10^2$ & $1.64$ \\\hline
\end{tabular}
\end{center}
\caption{The same as Table \ref{Drac} for the observation of Sagittarius. The first two columns related to EGRET and FERMI are current constraints whereas the last one corresponding to CTA is an estimation of its possible sensitivity.}
\label{Sag}
\end{table}

\begin{table}[t]
\begin{center}
\begin{tabular}{|c|c|c|c|}
\hline
\hline
\multicolumn{4}{|c|}{Canis Major dSph} \\
\hline
\hline
Experiment& EGRET & FERMI & CTA \\
\hline
$A_{Eff}$ & $1.5\times10^3$ & $10^4$ & $4\times10^8$ \\
\hline
$\Delta\Omega$ & $10^{-3}$  & $\sim10^{-5}$  & $10^{-5}$ \\
\hline
$t_{exp}$ &  \multicolumn{2}{|c|}{1\;yr} & 100 h \\
\hline
$\Phi_{Bg}$ &  \multicolumn{2}{|c|}{$3.87\times10^{-6}$} & $2.7\times10^{-4}$(**) \\
\hline
$\Phi_{\gamma}^{(5)}$  & $1.71\times10^{-6}$ & $1.08\times10^{-5}$ & $1.25\times10^{-8}$ \\
\hline
$\Phi_{\gamma}^{(2)}$ & $6.15\times10^{-7}$ & $2.94\times10^{-6}$ & $4.82\times10^{-9}$ \\
\hline
$\langle J\rangle_{\Delta\Omega}$ & 8.3 & \multicolumn{2}{|c|}{139.9} \\
\hline
$N_\gamma^{(5)}\langle\sigma v\rangle/M^2$  & $2.60\times10^3$ & $9.68\times10^2$ & $1.12$ \\
\hline
$N_\gamma^{(2)}\langle\sigma v\rangle/M^2$ & $9.32\times10^2$ & $2.64\times10^2$ & $0.43$ \\\hline
\end{tabular}
\end{center}
\caption{The same as Table \ref{Sag} for the observation of Canis Major.}
\label{CM}
\end{table}

On the other hand, we have assumed an Einasto profile for SEGUE 1 since it is more consistent with observations \cite{Mag11}:
\begin{equation}
\rho_{\text{Einasto}}(r)\,=\,\rho_s \text{exp}\left\{-2n\left[\left(\frac{r}{r_s}\right)^{1/n}-1\right]\right\}
\end{equation}
with the scale density $\rho_s=1.1\times10^8M_\odot \,\text{Kpc}^{-3}$, the scale radius $r_s=0.15\,\text{Kpc}$ and the index $n=3.3$ \cite{Mag11}.
Very recent observations \cite{Aleksic:2011qh} of SEGUE 1 (considered by many authors as possibly the most DM dominated satellite galaxy known in our galaxy) by MAGIC found  no significant gamma-ray emission above the background when taking into account the spectral features of the gamma-ray spectrum of specific DM models in a supersymmetric scenario.

In any case, both the modification of the density profile and the introduction of substructures just include an additional constant in
the analysis that  is easy to update. With our assumptions, the values of the astrophysical factor for each source are reported
in Tables %\ref{Drac}, \ref{Sag}, \ref{CM}, \ref{GC}, \ref{SEG}.
\ref{Drac}-\ref{SEG}.

%\subsection{Minimun detectable flux}
We can estimate the minimum detectable flux $\Phi_\gamma$ taking into account the total number of
observed gamma rays. Due to the uncertainties of these kind of analysis, it is usual to demand a significance of
at least $5\sigma$. For an observed target presenting: exposure time of $t_{exp}$ seconds,  instrument of effective area $A_{eff}$ and angular acceptance
$\Delta\Omega$, the significance of the detection exceeding $5\sigma$ (or alternatively $2\sigma$) is:
\begin{equation}
\frac{\Phi_\gamma\sqrt{\Delta\Omega\,A_{eff}\,t_{exp}}}{\sqrt{\Phi_\gamma+\Phi_{Bg}}} \geq 5\; (2).
\label{minflu}
\end{equation}

The DM annihilation flux $\Phi_\gamma$ and the background flux $\Phi_{Bg}$ are given in $\text{cm}^{-1}\text{s}^{-1}\text{sr}^{-1}$ \cite{Ev04}.
The evaluation of the background $\Phi_{Bg}$ and its value depends both on the experiment and on the source. In the case of satellite experiments, the diffuse gamma rays flux from astrophysical sources is the only contribution to the background depending on the location of the source \cite{Ev04}. The background for FERMI is assumed to be the same as for EGRET \cite{Berg08,webE,Ev04}:
\begin{eqnarray}
 \frac{\text{d}N_{Bg-a}}{\text{d}E}\approx \mathcal{N}\,\text{GeV}^{-1}\text{cm}^{-2}\text{s}^{-1}\text{sr}^{-1}\times
 \left(\frac{100\,\text{GeV}}{E_{Bg}}\right)^{2.1}\;,
 \label{bg}
\end{eqnarray}
and the exposition time $t_{exp}=1$ yr, common for the satellite experiments. We have chosen the spectral index $2.1$ since it is the most conservative value (see \cite{Ev04} or \cite{webE} for the particular values of $\mathcal{N}$ for the particular targets). In reality, the effective area of any detector depends on the particular energy at which operates. Eq. (\ref{minflu}) assumes a constant value and it is the main approximation of this
equation. We will assume a typical effective area for EGRET of $1.5\times10^3$ cm$^{2}$, whereas we will use $A_{eff}=10^4$ cm$^{2}$ for FERMI. On the contrary, the angular acceptance is much larger for EGRET: $\Delta\Omega=10^{-3}\,\text{sr}$, than for FERMI: $\Delta\Omega=10^{-5}\,\text{sr}$ \cite{EGRET,Fer,FER,Berg08}.

\begin{table}[t]
\begin{center}
\begin{tabular}{|c|c|c|c|}
\hline
\hline
\multicolumn{4}{|c|}{Galactic Center} \\
\hline
\hline
Experiment& EGRET & FERMI & CTA \\
\hline
$A_{Eff}$ & $1.5\times10^3$ & $10^4$ & $4\times10^8$ \\
\hline
$\Delta\Omega$ & $10^{-3}$  & $\sim10^{-5}$  & $10^{-5}$ \\
\hline
$t_{exp}$ &  \multicolumn{2}{|c|}{1\;yr} & 100 h \\
\hline
$\Phi_{Bg}$ &  \multicolumn{2}{|c|}{$1.2\times10^{-4}$} & $2.7\times10^{-4}$(**) \\
\hline
$\Phi_{\gamma}^{(5)}$  & $8.23\times10^{-6}$ & $3.51\times10^{-5}$ & $1.25\times10^{-8}$ \\
\hline
$\Phi_{\gamma}^{(2)}$ & $3.23\times10^{-6}$ & $1.30\times10^{-5}$ & $4.82\times10^{-9}$ \\
\hline
$\langle J\rangle_{\Delta\Omega}$ & 26 & \multicolumn{2}{|c|}{280} \\
\hline
$N_\gamma^{(5)}\langle\sigma v\rangle/M^2$  & $3.98\times10^3$ & $1.57\times10^3$ & $0.56$ \\
\hline
$N_\gamma^{(2)}\langle\sigma v\rangle/M^2$ & $1.56\times10^3$ & $5.83\times10^2$ & $0.21$ \\
\hline
\end{tabular}
\end{center}
\caption{The same as Table \ref{Sag} for the observation in the direction of the Galactic Center.}
\label{GC}
\end{table}

\begin{table}[t]
\begin{center}
\begin{tabular}{|c|c|c|}
\hline
\hline
\multicolumn{3}{|c|}{SEGUE 1} \\
\hline
\hline
Experiment& MAGIC-I & CTA\\
\hline
$A_{Eff}$ & $\sim10^8$ & $10^{10}$ \\
\hline
$\Delta\Omega$ & $10^{-5}$ & $10^{-5}$  \\
\hline
$t_{exp}$ & 29.4 h & 50 h \\
\hline
$\Phi_{Bg}$ & $1.9\times10^{-7}$(**) & $10^{-7}$\\
\hline
$\Phi_{\gamma}^{(5)}$  & $3.61\times10^{-7}$ &$1.25\times10^{-8}$ \\
\hline
$\Phi_{\gamma}^{(2)}$ & $1.06\times10^{-7}$ &$4.83\times10^{-9}$  \\
\hline
 $\langle J\rangle_{\Delta\Omega}$ &  \multicolumn{2}{|c|}{11.4} \\
\hline
$N_\gamma^{(5)}\langle\sigma v\rangle/M^2$  &$3.97\times10^2$ & $13.8$  \\
\hline
$N_\gamma^{(2)}\langle\sigma v\rangle/M^2$ & $1.16\times10^2$&$5.32$ \\
\hline
\end{tabular}
\end{center}
\caption{The same as Table \ref{Drac} for the observation of SEGUE 1. The first column  corresponds  to MAGIC-I and establishes the present constraint from this target whereas the second one estimates the CTA prospects. SEGUE 1 is the only target for which we have assumed an Einaisto profile for the DM distribution \cite{Mag11}.}
\label{SEG}
\end{table}

In the case of ground-based experiments, besides the above diffuse gamma-ray flux, there are two other sources of background: the hadronic and the cosmic-ray electrons. In any case, the hadronic source dominates at high energies for which the ground-based experiments are sensitive.
Taking into account the data observed by the Whipple 10 m telescope, it is possible to find a estimation for this background rate \cite{Berg08, BUB}:
\begin{eqnarray}
 \frac{\text{d}\Phi_{Bg-h}}{\text{d}E}\approx\epsilon\times10^{-5}\,\text{GeV}^{-1}\text{cm}^{-2}\text{s}^{-1}\text{sr}^{-1}\times
 \left(\frac{100\,\text{GeV}}{E_{Bg}}\right)^{2.7}\;,
 \label{bg}
\end{eqnarray}
 where we have integrated above the 100 GeV threshold of MAGIC,  estimating its effective area as  $A_{eff}=5\times10^{8}\,\text{cm}^{2}$ and its angular acceptance by $\Delta\Omega=10^{-5}\text{sr}$. %for Draco dSph ($t_{exp}=40h$) ),
The parameter  $\epsilon$ which corresponds to the fraction of hadronic shower which is misidentified as
electromagnetic  is set to the order of $1\%$ for MAGIC \cite{Mag11,Mag,Berg08}.

%\twocolumn

%
%
%

We are also interested in estimating the sensitivity of the next generation of ground-based VHE gamma-ray CTA instruments. Although there are still many details of the CTA project to be fixed,
an important improvement in the effective area is expected thanks to the large number of
telescopes in the array. We will assume a typical effective area of $A_{eff}\sim 1$ km$^2$, whereas the improvement in angular acceptance and background discrimination will be typically of order one \cite{cta}. In Tables \ref{Drac}-\ref{SEG}
%, \ref{Sag}, \ref{CM}, \ref{GC}, \ref{SEG},
we report the technical details of each experiment, the background estimations and the resulting values of the minimum detectable gamma-ray fluxes for the Draco, Sagittarius, Canis Major dSphs, for the Galactic Center and for SEGUE 1 respectively.

%The effective area depends on the zenith of the source with respect the telescope: we refer to J.Albert et al. \cite{Alb07} for the value of MAGIC-I
%effective area at 100 GeV, that is $\sim10^{8}$ for the SEGUE 1 observation ($t_{exp}=29.4h$).  An estimation of the background for SEGUE 1 observation was been calculated by means of the expression (\ref{bg}) \cite{Berg08}.\\

\begin{figure}[t]
\begin{center}
\epsfxsize=10cm   %width of figure - will enlarge/reduce the figures
%\epsfbox{ajustes_graphics_GLAST_pp1.eps}
%\figurebox{2cm}{3cm}{} %to have a box alone
\resizebox{10.8cm}{8.4cm} %$\resizebox{8.5cm}{!}
{\includegraphics{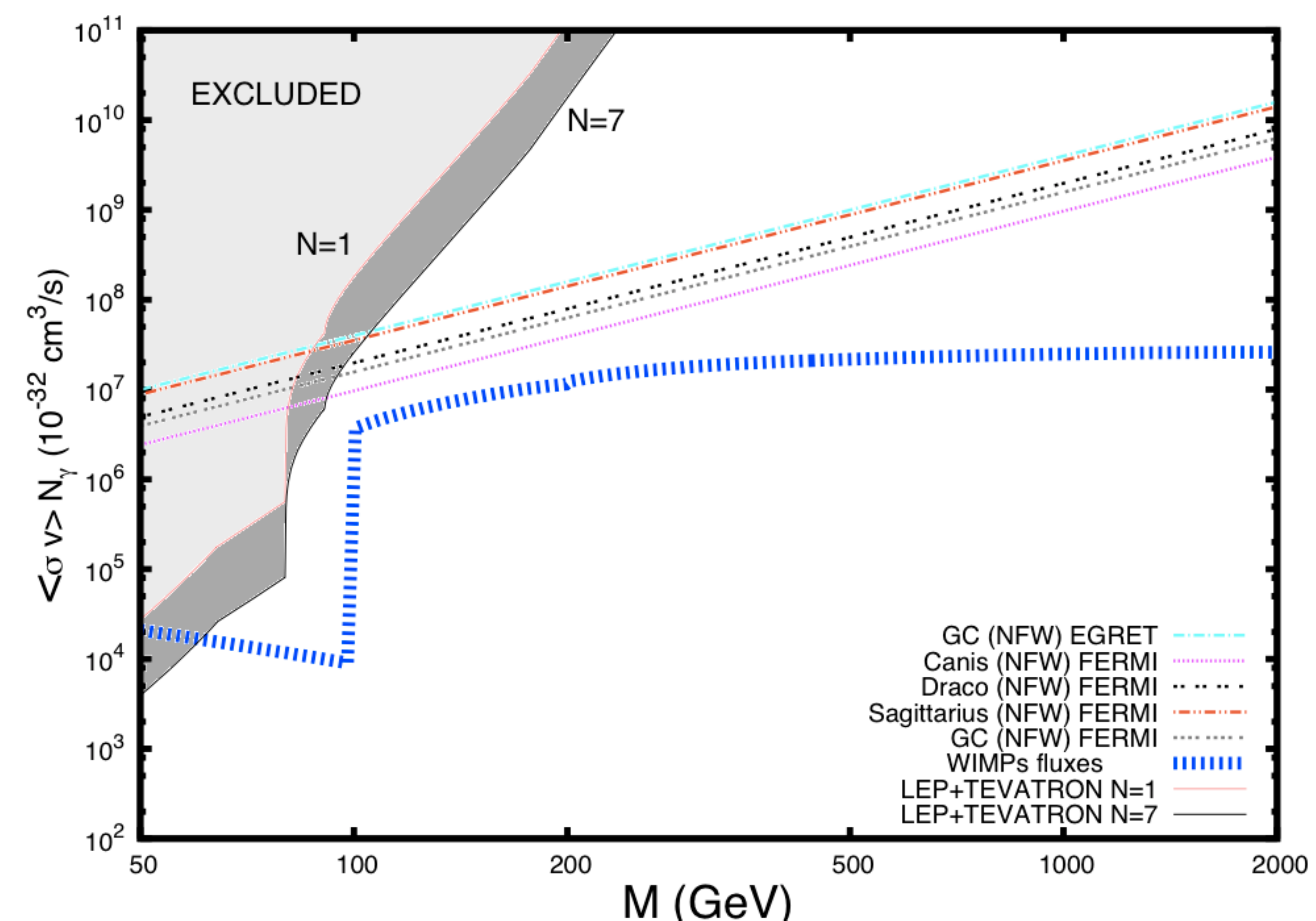}}
\caption {\footnotesize Sensitivity of different targets to constrain gamma rays coming from branon annihilation. The straight lines
show the estimated exclusion limits at $5\sigma$ for satellite experiments (FERMI and EGRET).
The blue thick line corresponds to the photon flux above 1 GeV coming from branons with the thermal abundance inside the WMAP7  \cite{WMAP} %\cite{Komatsu:2010fb}
limits ($\Omega_{\text{CDM}} h^2 = 0.1123\pm0.0035$).
The area on the upper left corner above the corresponding lines is excluded by $\text{LEP}$ and $\text{TEVATRON}$ experiments for both $N = 1$ and $N=7$, number of extra dimensions.}
\label{FER}
\end{center}
\end{figure}

\begin{figure}[t]
\begin{center}
\epsfxsize=10cm   %width of figure - will enlarge/reduce the figures
%\epsfbox{ajustes_graphics_GLAST_2p.eps}
%\figurebox{2cm}{3cm}{} %to have a box alone
\resizebox{10.8cm}{8.4cm}  %$\resizebox{8.5cm}{!}
{\includegraphics{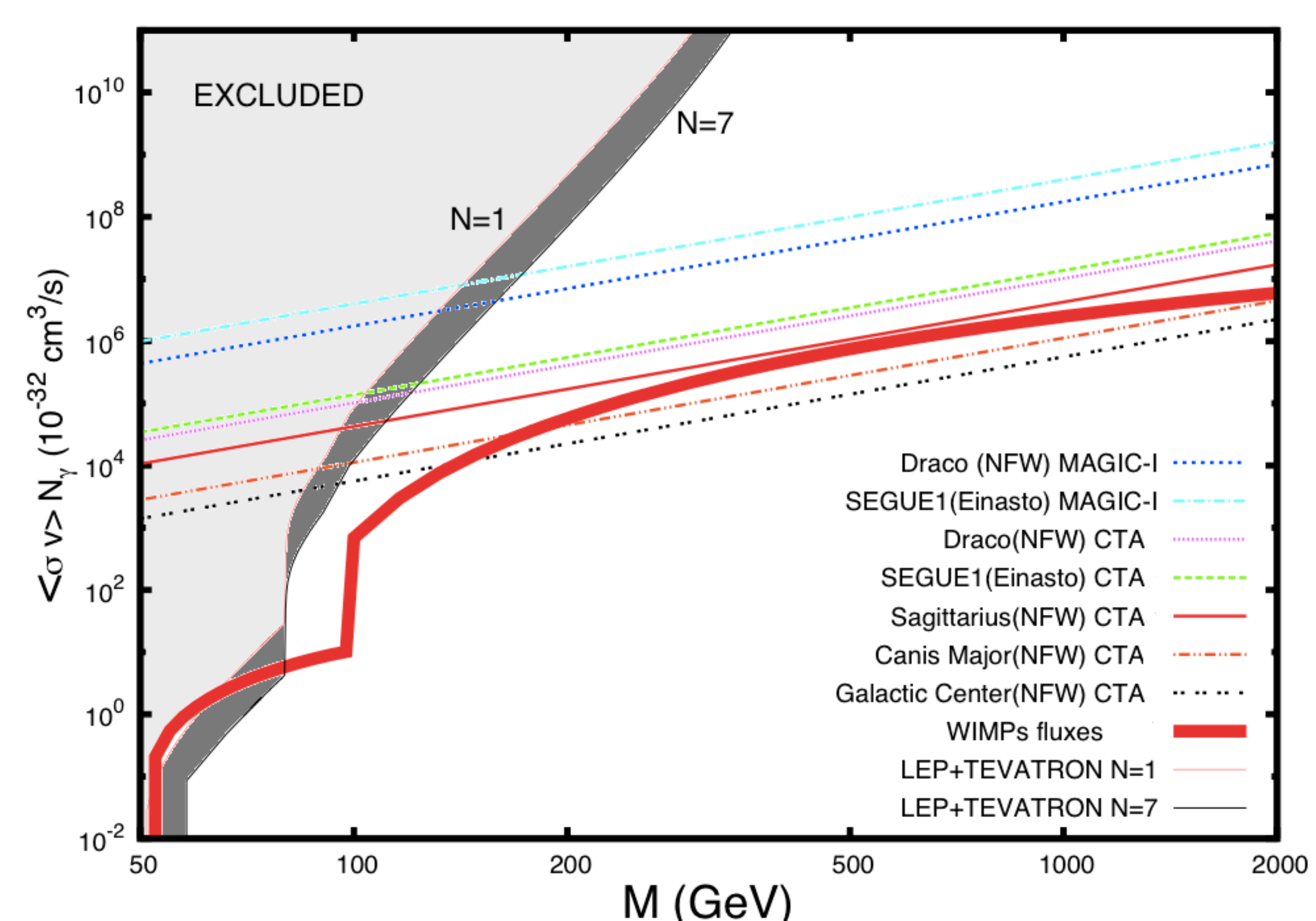}}
\caption {\footnotesize Same as Fig. \ref{FER} for ground-based detectors. In this figure, the continuous thick red line
corresponds to the photon flux above 50 GeV coming from branons with the thermal abundance inside the WMAP7 \cite{WMAP} limits.}
\label{ACT}
\end{center}
\end{figure}

By using the estimated minimum detectable flux at 2$\sigma$ or 5$\sigma$ significance and the particular astrophysical factor ($J_{\langle\Delta\Omega\rangle}$) of each target, the sensitivity on $N_\gamma^{(2,5)}
\langle\sigma v\rangle$   has been obtained as a function of the WIMP mass depending on the particular detector. The assumption of a constant effective area gives the typical power law behavior on the WIMP mass of this sensitivity  ($N_\gamma^{(2,5)}
\langle\sigma v\rangle\varpropto M^2$).  The corresponding curves for the different targets and
detectors are shown in Figs. 2 and 3. On the other hand, the theoretical value for  $N_\gamma
\langle\sigma v\rangle$ for branons has been obtained by integrating the differential
spectrum $\sum_i\langle\sigma_i v\rangle
\frac{\text{d}\,N_\gamma^i}{\text{d}\,E_{\gamma}}$ taking into account the energy threshold
of 1 GeV for satellite experiments (Fig. 2) or 50 GeV for ACTs (Fig. 3). The resulting  $N_\gamma
\langle\sigma v\rangle$ is a function of the two branon parameters $(f,M)$ and it does not depend on $N$ (number of branon species). 
This can be easily understood due to the fact that
the proportionally lower flux coming from the annihilation of a larger number of branon species, is compensated by the higher abundance that a larger number of species provides (for a fixed coupling, i.e. for a fixed value of $f$). Assuming that
the branon relic density agrees with WMAP observations \cite{WMAP}, it is possible to obtain $f(M)$, which finally allows us to plot  $N_\gamma\langle\sigma v\rangle$ as a function of the branon mass.
Thus, if the integrated spectrum line is over the straight lines  (which represent the sensitivity at $5\sigma$ for a particular target), a detector will be sensitive to  branon
annihilation coming from a particular target. We see that present experiments (EGRET, FERMI or MAGIC) are unable to
detect signals from branon annihilation for the targets considered. However, as shown in
Fig. 3, future experiments such as CTA
could be able to detect gamma-ray photons coming  from the annihilation of branons with masses
higher than 150 GeV for observations of the Galactic Center or above 200 GeV for Canis Major.

It is important to note again that the above computations and figures are based on particular assumptions about the DM profiles and neglecting substructure contributions. Uncertainties of order one are expected for dwarf spheroidals, but existence of boost factors of up to three orders of magnitude has been claimed for Galactic Center analyses \cite{Prada:2004pi}. Even in this case, after having taken these uncertainties into account, only satellite experiments may have already observed gamma rays from branon annihilation and mainly in the low range of the spectrum.

\section{Conclusions}

We have studied the sensitivity of different gamma-ray telescopes for the observation of indirect signals of branon dark matter in brane-world scenarios.
Under the assumption that branons are mass degenerate, this sensitivity only depends on two parameters of the
effective theory that describes the low energy dynamics of flexible brane-worlds: the brane tension scale $f$
and the branon mass $M$.

We have computed the production of photons coming from branon annihilation happening in either some
dSphs or the Galactic Center, and estimated the sensitivity for these cosmic photons to be detected in different experiments.
In particular, we have studied the prospective detectable flux from Draco, Sagittarius, Canis Major, SEGUE 1 and for the Galactic Center
for EGRET, FERMI and the future CTA. In the case of Draco and SEGUE 1, an estimation for the MAGIC telescope is given as well.
The estimated constraints show that the interesting parameter space of the theory, where the thermal branon relics account
for the total non-baryonic dark matter content of the Universe, has not been restricted by present observation yet.

%\subsection{Results}

Concerning the next generation of ACTs,
they seem to be able to prove this thermal area of the parameter space thanks to, fundamentally, the use of a large number of telescopes which can increase significantly  the
effective area of detection (Fig. \ref{ACT}). With a better sensitivity, this kind of instruments could explore the highest part of the spectrum for branons heavier than 200 GeV by means of the observation of both dSphs (Canis Major in particular) and the Galactic Center.

Therefore, the technical instrumentation available nowadays appears to be insufficient in order to constrain the brane-world models through gamma-ray detection. On the other
hand, the estimates for the next generation of ACTs show that these types of signals could provide the first evidences of these models. In the same figures
(\ref{flux} and \ref{ACT}), it is possible to see the present constraints from collider experiments. These searches
are complementary and prove, in general, a different area of the parameter space of the model.

%%%%%%%%%%%%%%%%%%%%%%%%%%%%%%%%%%%%%

\begin{acknowledgments}
We would like to thank Daniel Nieto for useful comments.
This work has been supported
by MICINN (Spain) project numbers FIS 2008-01323,
FPA 2008-00592 and Consolider-Ingenio MULTIDARK
CSD2009-00064. AdlCD also acknowledges financial
support from URC research fellowships and National Research Foundation (South Africa) 
and kind hospitality of UCM, Madrid
while elaborating part of the manuscript.
\end{acknowledgments}

%%%%%%%%%%%%%%%%%%%%%%%%%%%%%%%%%%%%%%%%%%

\end{document}